\documentclass[nopacs,epj]{svjour}
\usepackage{graphicx}

\usepackage{amssymb}
\usepackage{amsmath}

\usepackage{multirow}

\usepackage{color}

\usepackage[colorlinks]{hyperref}

\AtBeginDocument{%
    \hypersetup{
      urlcolor     = blue, 
      linkcolor    = black, 
      citecolor   = blue, 
      pdfborder={0 0 0},
      urlbordercolor={1 1 1},
      citebordercolor={1 1 1}
    }
    }

\begin{document}

\hyphenation{coun-ter cor-res-pon-din-gly e-xam-ple
co-o-pe-ra-tion ex-pe-ri-men-tal pum-ping o-pe-ra-tors in-dis-tin-gui-sha-bi-li-ty}

\title{An example of quantum imaging: rendering an object undetectable}

\titlerunning{An example of quantum imaging: rendering an object undetectable}

\authorrunning{Stefan Ataman}

\author{Stefan~Ataman\inst{1}
}                     

\institute{Extreme Light Infrastructure - Nuclear Physics
(ELI-NP), 30 Reactorului Street, 077125 Magurele, 
Romania \email{stefan.ataman@eli-np.ro}}


%
%
%
\date{Received: date / Revised version: date}
%
\abstract{In this paper we propose and analyse a
Gedankenexperiment involving three non-linear crystals and two
objects inserted in the idler beams. We show that, besides the behaviour that can be
extrapolated from previous experiments involving two crystals and
one object, we are able to predict a new effect: under certain
circumstances, one of the objects can be rendered undetectable to any
single detection rate on the signal photons with discarded idler
photons. This effect could find
applications in future developments of quantum imaging techniques.
%
\PACS{
      {PACS-key}{describing text of that key}   \and
      {PACS-key}{describing text of that key}
     } 
} 
%


\maketitle
%

\section{Introduction}

The non-linear optical process of spontaneous parametric
down-conversion (SPDC) \cite{Bur70,Kly67,Hon85} creates pairs of
highly entangled single photon pairs \cite{Rub94-Gri97-Kel97,Kwi95b}. Its use
became ubiquitous in quantum optics (QO), where numerous
experiments rely on this process
\cite{HOM87,Pit96,Ou90F,Ou90a,Ou90b,Zou91,Lem14,Heu15}. 

SPDC is a
process that happens randomly, somehow in the spirit of
spontaneous emission of an excited atom. However, in an experiment
by Ou, Wang, Zou and Mandel \cite{Ou90a,Ou90b}, two non-linear
crystals fed by the same pump laser show a phenomenon of ``phase
memory''.
%
%
%
A landmark experiment using two  non-linear crystals was performed
by Zou, Wang and Mandel \cite{Zou91}. Its counter-intuitive
feature stemmed from the fact that the singles count rate of the
two interfering signal beams shows or does not show interference
effects in function of the distinguishability/indistinguishability
of the discarded idler beams. Based on this principle, the imaging
of small objects using undetected (discarded) photons has been
recently experimentally demonstrated \cite{Lem14}. One can now use
a wavelength where the object is transparent and detect photons of
another wavelength, more convenient for the detectors
\cite{Lem14}. This idea is pushed even further out of the lab and towards industrial applications by Kalashnikov et
al. in reference \cite{Kal16} where they perform infrared spectroscopy yet they use red photons at the detectors. For a review on quantum imaging
techniques and implementations one can consult reference
\cite{Sim14}.


With these latest developments, it becomes obvious that the field of quantum imaging is still in its infancy and many developments are to be expected in the near future. 

In this paper, we show that an object can be rendered undetectable
using the already proven principle of quantum imaging with
discarded idler photons \cite{Zou91,Lem14}. We propose the
extension of the experiment from reference \cite{Zou91} to the
case involving three non-linear crystals. Besides the expected
behaviour (i.e. the extension of the counter-intuitive features of
the results of Zou et al. \cite{Zou91}), we show that, under
certain circumstances, the second object can be rendered invisible
to any photo-counters if the idler photons are discarded.



This paper is organized as follows. In Section
\ref{sec:the_experiment_three_crystals} the proposed experiment is
introduced and qualitatively discussed. Its complete quantum
optical description is thoroughly developed in Section
\ref{sec:quantum_optical_description_exp}. The computation of
various single and coincidence photo-detection rates is done in
Section \ref{sec:computation_singles_coinc_prob}. Finally,
conclusions are drawn in Section \ref{sec:conclusions}.


\section{The proposed experiment}
\label{sec:the_experiment_three_crystals} 
\begin{figure}
\centering
\includegraphics[width=3.3in]{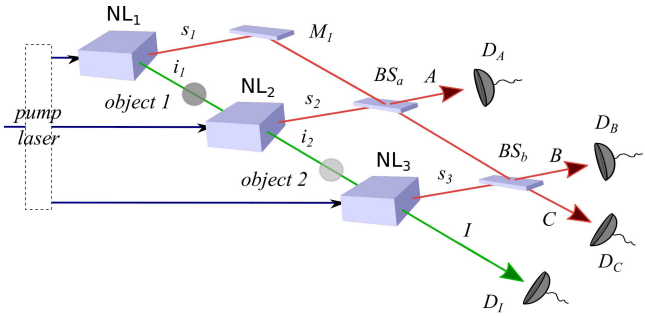}
\caption{The proposed experiment. Three non-linear crystals are
pumped by a common laser. The idler modes of the crystals overlap,
therefore become indistinguishable. The signal modes are brought
to interfere through the beam splitters $\text{BS}_a$ and
$\text{BS}_b$. The two objects can be inserted in the gray
regions of $i_1$ and $i_2$.} \label{fig:three_crystal_exp}
\end{figure}

In the proposed
Gedankenexperiment three non-linear crystals are pumped together
by a common laser (see Fig.~\ref{fig:three_crystal_exp}). We assume that the pumping power of each individual crystal can be adjusted at will. The
crystals are aligned in such a way, that the idler mode of $NL_1$
($i_1$) is overlapping the idler modes of $NL_2$ ($i_2$) and
$NL_3$ ($i_3$). Therefore, a photo-detection at detector $D_I$
would not be able to tell from which nonlinear crystal the photon
originated. The signal modes of $NL_1$ ($s_1$) and $NL_2$ ($s_2$)
are brought together into the beam splitter $\text{BS}_a$. Its
first output leads to the photo-detector $D_{A}$, while the second
one enters the beam splitter $\text{BS}_b$, where it combines with
the signal mode of $NL_3$ ($s_3$). The two outputs of this beam
splitter lead to the photo-detectors $D_{B}$ and $D_{C}$.
Throughout this paper, we assume ideal and broad-band
photo-detectors.

More or less transparent objects can be inserted in the dashed
regions of the beams $i_1$ and $i_2$ (see
Fig.~\ref{fig:three_crystal_exp}).

If we do not pump the non-linear crystal $NL_3$ and consider
single rates at the detector $D_A$, we find ourselves in the
experimental setup described by Zou, Wang and Mandel \cite{Zou91}.
Therefore, the same counter-intuitive behavior is expected, with
the singles detection rate at $D_A$ depending on the
transmissivity of the inserted object (called ``object 1'') into
the beam $i_1$.

At a first glance, the extension to three crystals and two objects
seems trivial and devoid of surprises. Inserting an object (called
thereafter ``object 2'') into the beam $i_2$ should be detected by
a variation of the singles rate at either $D_B$ or $D_C$.

In the following, we will show that this is not always the case.
For some well chosen parameters of object 1, beam splitter
$\text{BS}_a$ and the pumping powers of $NL_1$ and $NL_2$, object
2 can simply become invisible (regardless of its transmissivity)
to any single detection rate at $D_A$, $D_B$ or $D_C$. Therefore,
if we discard the idler photons, there is no way to detect object
2. Even coincidence counts $D_B-D_I$ and $D_C-D_I$ fail to reveal
its presence.


\section{The quantum optical description of the experiment}
\label{sec:quantum_optical_description_exp} In the SPDC process,
the input beam (generally called ``pump'', $p$) is split into two
output beams (generally called  ``signal'' $s$ and ``idler'' $i$).
This process takes place at the single photon level: one input
photon is destroyed and two output photons are created. Energy
($\hbar\omega$) and momentum ($\hbar\boldsymbol{k}$) conservation
requirements (also called ``phase matching conditions'') impose
frequency ($\omega_p=\omega_s+\omega_i$) and wavenumber
($\boldsymbol{k}_p\approx\boldsymbol{k}_s+\boldsymbol{k}_i$)
relations between the input-output fields.

We shall make a number of simplifying assumptions. Similar to
references \cite{Zou91,Lem14,Heu15,Wis00} we shall consider all
modes as being monochromatic. Since the phase relation between the
three pump fields can be fixed at any value, we simply assume that
all crystals are pumped in phase, yielding
\begin{eqnarray}
\label{eq:psi_input_a_p_plus_void}
\vert\psi_{\text{in}}\rangle=\frac{1}{\mathcal{N}}\big(\gamma_1\vert1_{p_1}\rangle
+\gamma_2\vert1_{p_2}\rangle+\gamma_3\vert1_{p_3}\rangle\big)
\nonumber\\
=\frac{\left(\gamma_1\hat{a}_{p_1}^\dagger+\gamma_2\hat{a}_{p_2}^\dagger
+\gamma_3\hat{a}_{p_3}^\dagger\right)}{\mathcal{N}}\vert0\rangle
\end{eqnarray}
where $\mathcal{N}$ is a normalization constant,
$\vert1_{p_k}\rangle$ denotes a Fock state with one photon in mode
(port) $p_k$, $\hat{a}_{p_k}^\dagger$ denotes the usual input
creation operator in mode $p_k$ with $k=1,2,3$  and
$\vert0\rangle$ is the vacuum state. A detailed discussion of this
choice is given in Appendix \ref{app:coherent_to_Fock_input}. The
$\gamma_k$ parameters with $k=1,2,3$ (assumed real and positive
throughout this paper) are connected to input power differences for identical non-linear crystals \cite{Her94,Suz08}. The quantum state
\eqref{eq:psi_input_a_p_plus_void} also implies that no two
down-conversions happen at the same moment. This is consistent
with the low down-conversion efficiency \cite{Ou90a,Zou91,Heu15} of the non-linear crystals.

From the perspective of QO, the field operator transformation in
each SPDC process is
\begin{equation}
\label{eq:SPDC_field_transf}
\hat{a}_{p_n}^\dagger\rightarrow\hat{a}_{s_n}^\dagger\hat{a}_{i_n}^\dagger
\end{equation}
where $\hat{a}_{s_n}^\dagger$ ($\hat{a}_{i_n}^\dagger$) denotes
the signal (idler) creation operators and $n=1,2,3$. The process
of SPDC creates pairs of photons from a pump photon in a
spontaneous, hence unpredictable way. However, in our model
\eqref{eq:psi_input_a_p_plus_void} a fixed phase relationship is
assumed among the parametric down-conversions (and set to zero for
convenience).

\begin{figure}
\centering
\includegraphics[width=3.3in]{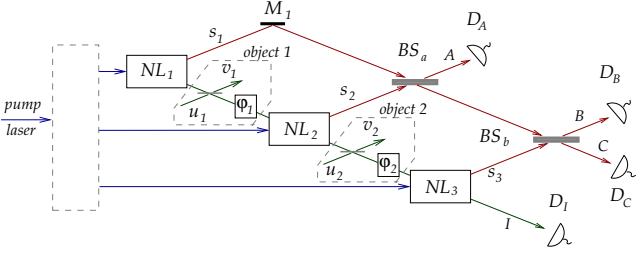}
\caption{The objects inserted in the idler beams of $NL_1$ and,
respectively, $NL_2$ can be modelled by two beam splitters having
transmissivity (reflectivity) coefficient $T_k$ ($R_k$) and two
phase delays $\varphi_k$ with $k=1,2$.}
\label{fig:three_crystal_exp_objects_as_BS}
\end{figure}

With the two objects inserted, the experiment is depicted in
Fig.~\ref{fig:three_crystal_exp_objects_as_BS}. Each object is
modelled through a beam splitter \cite{Leo03} having transmission
(reflection) coefficients $T_m$ ($R_m$) and a phase shift
$\varphi_m$ with $m=1,2$. Typical BS unitarity constrains
\cite{Lou03} apply to these coefficients. The inputs $u_m$ are
always in the vacuum mode while the fictitious modes $v_m$ have to
be considered, for reasons of unitary evolution of the system
(although in the final computation they will be traced out). For a
perfectly transparent (opaque) object $m$ we have $T_m=1$
($T_m=0$) with $m=1,2$.

For simplicity, in the main part of the paper, the optical lengths
of $s_1$ and $s_2$ leading to $\text{BS}_a$ are assumed to be
equal. The same assumption is taken for $s_2$ and $s_3$. The
scenario with unequal lengths in the signal beams is discussed in
Appendix \ref{sec:appendix_field_operator_transf}.

Beam splitters $\text{BS}_a$ and $\text{BS}_b$ are characterized
by the transmission (reflection) coefficients $T_n$ ($R_n$) with
$n=\{a,b\}$. Unitarity requirements impose the well-known
conditions $\vert{T_n}\vert^2+\vert{R_n}\vert^2=1$ and
$T_nR_n^*+T_n^*R_n=0$ \cite{Lou03}. Whenever we shall assume
balanced beam splitters, we shall replace $T_n=1/\sqrt{2}$ and
$R_n=i/\sqrt{2}$.

After a series of calculations (see details for a slightly more
general case in Appendix
\ref{sec:appendix_field_operator_transf}), the output state vector
is found to be
\begin{eqnarray}
\label{eq:psi_output}
\vert\psi_{\text{out}}\rangle=\frac{1}{\mathcal{N}}\bigg(T_2\text{e}^{i\varphi_2}\left(\gamma_1R_aT_1\text{e}^{i\varphi_1}
+\gamma_2T_a\right)\vert1_{A}1_I\rangle
\nonumber\\
+R_2\left(\gamma_1R_aT_1\text{e}^{i\varphi_1}
+\gamma_2T_a\right)\vert1_{A}1_{v_2}\rangle
\nonumber\\
+\gamma_1R_aR_1\vert1_{A}1_{v_1}\rangle
\nonumber\\
+\left(R_bT_2\text{e}^{i\varphi_2}\left(\gamma_1T_aT_1\text{e}^{i\varphi_1}+\gamma_2R_a\right)+\gamma_3T_b\right)\vert1_{B}1_I\rangle
\nonumber\\
+R_bR_2\left(\gamma_1T_aT_1\text{e}^{i\varphi_1}+\gamma_2R_a\right)\vert1_{B}1_{v_2}\rangle
\nonumber\\
+\gamma_1T_aR_bR_1\vert1_{B}1_{v_1}\rangle
\nonumber\\
+\left(T_bT_2\text{e}^{i\varphi_2}\left(\gamma_1T_aT_1\text{e}^{i\varphi_1}+\gamma_2R_a\right)+\gamma_3R_b\right)\vert1_{C}1_I\rangle
\nonumber\\
+T_bR_2\left(\gamma_1T_aT_1\text{e}^{i\varphi_1}+\gamma_2R_a\right)\vert1_{C}1_{v_2}\rangle
\nonumber\\
+\gamma_1T_aT_bR_1\vert1_{C}1_{v_1}\rangle\bigg)\:
\end{eqnarray}
where $\vert1_{A}1_{I}\rangle$ denotes a Fock state with one
photon in mode (port) $A$ and one in mode $I$ etc. We construct
the density matrix
$\hat{\rho}_{\text{out}}=\vert\psi_{\text{out}}\rangle\langle\psi_{\text{out}}\vert$
and trace it out over the fictitious modes $v_1$ and $v_2$,
yielding the reduced density matrix
\begin{eqnarray}
\label{eq:rho_traced_v1_v2}
\hat{\rho}_{\text{red}}=\text{Tr}_{v_1,v_2}\left\{\hat{\rho}_{\text{out}}\right\}
=\sum_{m,\:n}{\langle{m_{v_1}n_{v_2}}\vert\hat{\rho}_{\text{out}}\vert{m_{v_1}n_{v_2}}\rangle}
\end{eqnarray}

\section{Singles and coincidence rates evaluation}
\label{sec:computation_singles_coinc_prob} Having the reduced
density matrix \eqref{eq:rho_traced_v1_v2} allows one to make all
predictions regarding the probabilities of singles or coincidence
detections. For example, the singles detection probability at the
photo-detector $D_{A}$ is given by
\begin{eqnarray}
P_{A}=\text{Tr}\left\{\hat{a}_{A}^\dagger\hat{a}_{A}\hat{\rho}_{\text{red}}\right\}
\end{eqnarray}
and after a short computation one arrives at
\begin{eqnarray}
\label{eq:P_s1_rho_traced_v1_v2} P_{A}\sim \vert
\gamma_1R_aT_1\text{e}^{i\varphi_1}+\gamma_2T_a\big\vert^2+\gamma_1^2\vert{R_a}R_1\vert^2
\end{eqnarray}
It is noteworthy to mention that the dependence on $T_2$ ($R_2$)
disappears during this computation, therefore the final result
from equation \eqref{eq:P_s1_rho_traced_v1_v2}, as expected, does
not depend on the properties of object $2$. If one assumes equal
pumping powers to $NL_1$ and $NL_2$ (i.e. $\gamma_1=\gamma_2$) and
a balanced beam splitter $\text{BS}_a$, equation
\eqref{eq:P_s1_rho_traced_v1_v2} simplifies to
\begin{eqnarray}
\label{eq:P_s1_final_balanced_BS} P_{A}\sim
1-T_1\sin\left(\varphi_1\right)
\end{eqnarray}
where we assumed $T_1$ to be real. This result is similar to the
ones already reported \cite{Zou91,Lem14}. Its paradoxical nature
arises from the fact that the two idler modes ($i_1$ and $i_2$) do
not participate in the interference of the signal modes ($s_1$ and
$s_2$). However, their indistinguishability/distinguishability
allows one to have or not to have a quantum state that can be
factorized. It is worthwhile to note that the
distinguishability/indistinguishability of the beams $i_2$ and
$i_3$ at the detector $D_I$ plays no role in measuring the
probability of detection at $P_{A}$.

The singles detection probability at $D_{B}$ is given by
$P_{B}=\text{Tr}\left\{\hat{a}_{B}^\dagger\hat{a}_{B}\hat{\rho}_{\text{red}}\right\}$
and one obtains
\begin{eqnarray}
\label{eq:P_s2_rho_traced_v1_v2} P_{B}\sim \big\vert
R_bT_2\text{e}^{i\varphi_2}\left(\gamma_1T_aT_1\text{e}^{i\varphi_1}+\gamma_2R_a\right)+\gamma_3T_b\big\vert^2
\nonumber\\
+\vert{R_bR_2}\left(\gamma_1T_aT_1\text{e}^{i\varphi_1}+\gamma_2R_a\right)\vert^2+\gamma_1^2\vert{T_a}R_bR_1\vert^2\:
\end{eqnarray}
At first sight, this result is simply the extension of equation
\eqref{eq:P_s1_rho_traced_v1_v2} to the case involving three
crystals and two objects. However, one can select values of
$\gamma_1$, $\gamma_2$, $T_a$, $T_1$ and $\varphi_1$ so that
\begin{equation}
\label{eq:condition_for_object2_invisible}
\gamma_1T_aT_1\text{e}^{i\varphi_1}+\gamma_2R_a=0
\end{equation}
In this case, equation \eqref{eq:P_s2_rho_traced_v1_v2} simplifies
to
\begin{eqnarray}
\label{eq:P_s2_obj_1_blocks}
P_{B}\sim\gamma_1^2\vert{T_a}R_bR_1\vert^2+\gamma_3^2\vert{T_b}\vert^2
\end{eqnarray}
In other words, the presence or absence of object $2$ has no
influence on the singles photo-count result at $D_{B}$. A similar
result is obtained at the detector $D_{C}$, namely
\begin{equation}
P_{C}\sim\gamma_1^2\vert{T_a}T_bR_1\vert^2+\gamma_3^2\vert{R_b}\vert^2.
\end{equation}
Even more counter-intuitive is the coincidence count result at
detectors $D_{B}$ and $D_I$, yielding
\begin{eqnarray}
\label{eq:Pc_s2_i_obj_1_blocks}
P_{B-I}\sim\gamma_3^2\vert{T_b}\vert^2
\end{eqnarray}
or, similarly, at detectors $D_{C}$ and $D_I$, giving
\begin{eqnarray}
P_{C-I}\sim\gamma_3^2\vert{R_b}\vert^2
\end{eqnarray}

The question arises: did object $2$ become completely undetectable
if we obey the condition from equation \eqref{eq:condition_for_object2_invisible}?
The answer is ``almost''. With the current configuration, we could
probe the presence of object $2$ by measuring the single counts at
$D_I$ or the coincidence counts at $D_{A}$ and $D_I$. The latter,
for example, yields
\begin{eqnarray}
\label{eq:Pc_s1_i_obj_1_blocks}
P_{A-I}
\sim\gamma_1^2\vert{T_1T_2}\vert^2/\vert{R_a}\vert^2
\end{eqnarray}
a result depending, indeed, on $T_2$. However, if we discard the
idler photons (as done in reference \cite{Lem14}), object 2 is
completely invisible to any single detection rates.

\section{Conclusions}
\label{sec:conclusions} In this paper we proposed and discussed an
experiment able to render undetectable -- using quantum, rather
than classical principles -- a more or less transparent object. For
well chosen experimental parameters (including only the first
object), a condition for the undetectability of the second object
can be obtained. In this case, no matter what properties the
second object has, its presence or absence will remain undetected
to all single detection rates if the idler photons are discarded.

\section*{Acknowledgments}
The author wishes to thank Dr. Radu Ionicioiu for useful
suggestions for the final form of the manuscript and Dr. Waleed
Mouhali for helping him double-check the main results of this
paper. The author also wishes to thank the anonymous reviewers for
their help in improving this paper.

\appendix

\section{About the input state of the system}
\label{app:coherent_to_Fock_input} Equation
\eqref{eq:psi_input_a_p_plus_void} might seem to be awkward, since
the inputs of the non-linear crystals are actually intense
coherent fields, while we consider them as being single-photon
Fock states. In order to justify this choice, we first write the
input state as a coherent state $\vert\alpha_{p_n}\rangle$ with
$n=1,2,3$ and expanding it in Fock state yields
\begin{equation}
\label{eq:pump_coherent_in_Fock}
\vert\alpha_{p_n}\rangle=e^{-\vert\alpha_{p_n}\vert^2/2}\left(\vert0\rangle
+\alpha_{p_n}\hat{a}_{p_n}^\dagger\vert0\rangle+\frac{\left(\alpha_{p_n}\hat{a}_{p_n}^\dagger\right)^2}{2}\vert0\rangle+\ldots\right)
\end{equation}
The only term from equation \eqref{eq:pump_coherent_in_Fock} that
yields two photons after the the down-conversion process
\eqref{eq:SPDC_field_transf} is the second one. This
down-conversion process has a (typically low) efficiency $g_n$
where this coefficient is connected to the $\chi^{(2)}$
nonlinearity of each crystal. Therefore, if we denote
$\gamma_n=\vert\alpha_n\vert g_n$ and assume post-selection of the
two-photon state only (higher order nonlinear processes are
negligible) takes us to the input state given by equation
\eqref{eq:psi_input_a_p_plus_void}. If one wants to avoid any higher-order nonlinear process in the crystals, number-resolving photo-detectors (e. g. superconducting nanowire \cite{Div08}) can be used instead of the usual ones.   

A similar result for the output state vector can be obtained by
considering the signal and idler modes initially in the vacuum
mode and by applying a parametric interaction Hamiltionian
\cite{Zou91,Sim14}.

\section{Computing the field operator transformations}
\label{sec:appendix_field_operator_transf}

\begin{figure}
\centering
\includegraphics[width=3.3in]{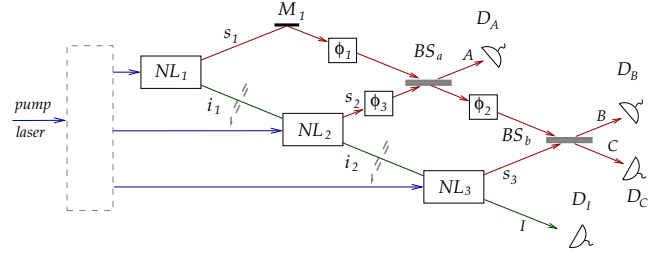}
\caption{The experiment in the general case. Path length
differences are modelled through phase delays $\phi_1$, $\phi_2$
and $\phi_3$, as depicted above.}
\label{fig:3crystal_experiment_general}
\end{figure}

Contrary to Section \ref{sec:quantum_optical_description_exp}, we
assume the lengths of the signal beams to the beam splitters
non-equal. Therefore, for example, the path length difference
(delay) in the beams $s_1$ and $s_2$ is modelled by the phase
shift $\phi_1$. Additional delays can be added, as depicted in
Fig.~\ref{fig:3crystal_experiment_general}.

\begin{figure}
\centering
\includegraphics[width=3.3in]{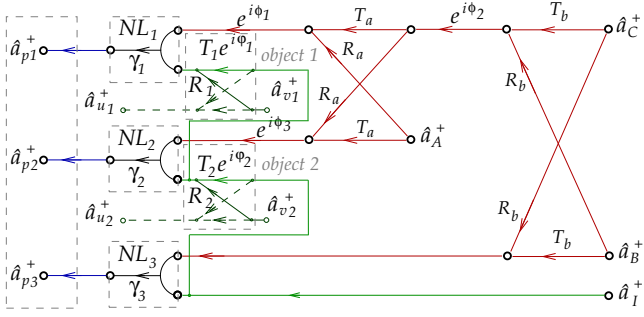}
\caption{The proposed Gedankenexperiment described with the
graphical method. We are interested in finding the input-output
(creation) field operator transformations
$\hat{a}_{p_k}^\dagger=f_k\left(\hat{a}_{A}^\dagger,\hat{a}_{B}^\dagger,\hat{a}_{C}^\dagger,\hat{a}_{I}^\dagger\right)$
i.e. the (operator) functions $f_k$ with $k=1,2,3$.}
\label{fig:graph_description_experiment}
\end{figure}

The singles/coincidence detection rates we are interested in can
be obtained in several ways. One could follow the approach from
reference \cite{Heu15} or the one from reference \cite{Wis00}.
Here, we sketch another way of computing them, using a graphical
method introduced in \cite{Ata14b} and extended to non-linear
optics in \cite{Ata15a}. We can draw the graph associated to our
experiment as depicted in
Fig.~\ref{fig:graph_description_experiment} and one obtains the
field operator transformations
\begin{eqnarray}
\label{eq:ap1_dagger_versus_output}
\hat{a}_{p_1}^\dagger=\gamma_1\left(R_a\text{e}^{i\phi_1}\hat{a}_{A}^\dagger
+T_aR_b\text{e}^{i\left(\phi_1+\phi_2\right)}\hat{a}_{B}^\dagger
+T_aT_b\text{e}^{i\left(\phi_1+\phi_2\right)}\hat{a}_{C}^\dagger\right)
\nonumber
\\
\times
\left(T_1\text{e}^{i\varphi_1}T_2\text{e}^{i\varphi_2}\hat{a}_{I}^\dagger
+T_1\text{e}^{i\varphi_1}\hat{a}_{v_2}^\dagger+R_1\hat{a}_{v_1}^\dagger\right)\quad
\end{eqnarray}

\begin{eqnarray}
\label{eq:ap2_dagger_versus_output}
\hat{a}_{p_2}^\dagger=\gamma_2\left(T_a\text{e}^{i\phi_3}\hat{a}_{A}^\dagger
+R_aR_b\text{e}^{i\left(\phi_2+\phi_3\right)}\hat{a}_{B}^\dagger
+R_aT_b\text{e}^{i\left(\phi_2+\phi_3\right)}\hat{a}_{C}^\dagger\right)
\nonumber
\\
\times
\left(T_2\text{e}^{i\varphi_2}\hat{a}_{I}^\dagger+R_2\hat{a}_{v_2}^\dagger\right)\quad
\end{eqnarray}

\begin{eqnarray}
\label{eq:ap3_dagger_versus_output}
\hat{a}_{p_3}^\dagger=\gamma_3\left(T_b\hat{a}_{B}^\dagger+R_b\hat{a}_{C}^\dagger\right)\hat{a}_{I}^\dagger
\end{eqnarray}
Starting from the input state vector
\eqref{eq:psi_input_a_p_plus_void} and applying the field operator
transformations
\eqref{eq:ap1_dagger_versus_output}--\eqref{eq:ap3_dagger_versus_output},
after some straightforward algebra one gets the output state
vector
\begin{eqnarray}
\label{eq:psi_output_GENERAL}
\vert\psi_{\text{out}}\rangle=\frac{1}{\mathcal{N}}\bigg(T_2\text{e}^{i\varphi_2}\left(\gamma_1R_aT_1\text{e}^{i\phi_1}\text{e}^{i\varphi_1}
+\gamma_2T_a\text{e}^{i\phi_3}\right)\vert1_{A}1_I\rangle
\nonumber\\
+\left(\gamma_1R_aT_1R_2\text{e}^{i\varphi_1}\text{e}^{i\phi_1}
+\gamma_2T_aR_2\text{e}^{i\phi_3}\right)\vert1_{A}1_{v_2}\rangle
\nonumber\\
+\gamma_1R_aR_1\text{e}^{i\phi_1}\vert1_{A}1_{v_1}\rangle
\nonumber\\
+\left(R_bT_2\text{e}^{i\phi_2}\text{e}^{i\varphi_2}\left(\gamma_1TT_1\text{e}^{i\phi_1}\text{e}^{i\varphi_1}
+\gamma_2R_a\text{e}^{i\phi_3}\right)+\gamma_3T_b\right)\vert1_{B}1_I\rangle
\nonumber\\
+R_bR_2\text{e}^{i\phi_2}\left(\gamma_1T_aT_1\text{e}^{i\phi_1}\text{e}^{i\varphi_1}
+\gamma_2R_a\text{e}^{i\phi_3}\right)\vert1_{B}1_{v_2}\rangle
\nonumber\\
+\gamma_1T_aR_bR_1\text{e}^{i\phi_1}\text{e}^{i\phi_2}\vert1_{B}1_{v_1}\rangle
\nonumber\\
+\left(T_bT_2\text{e}^{i\phi_2}\text{e}^{i\varphi_2}\left(\gamma_1T_aT_1\text{e}^{i\phi_1}\text{e}^{i\varphi_1}
+\gamma_2R_a\text{e}^{i\phi_3}\right)+\gamma_3R_b\right)
\vert1_{C}1_I\rangle
\nonumber\\
+T_bR_2\text{e}^{i\phi_2}\left(\gamma_1T_aT_1\text{e}^{i\phi_1}\text{e}^{i\varphi_1}+\gamma_2R_a\text{e}^{i\phi_1}\right)\vert1_{C}1_{v_2}\rangle
\nonumber\\
+\gamma_1T_aT_bR_1\text{e}^{i\left(\phi_1+\phi_2\right)}\vert1_{C}1_{v_1}\rangle\bigg)
\qquad
\end{eqnarray}
where $\mathcal{N}$ is a normalization constant. By imposing
$\phi_1=\phi_2=\phi_3=0$ equation \eqref{eq:psi_output_GENERAL}
migrates into equation \eqref{eq:psi_output} from Section
\ref{sec:quantum_optical_description_exp}.

\section{Generalization of the previous results}
\label{sec:appendix_general_case} We generalize here the results
from  Section \ref{sec:computation_singles_coinc_prob}. Using the
output wavevector from equation \eqref{eq:psi_output_GENERAL} and
following the same approach from Section
\ref{sec:computation_singles_coinc_prob}, the singles detection
probability at detector $D_B$ is found to be
\begin{eqnarray}
\label{eq:P_B_rho_traced_v1_v2_GENERAL} P_{B}\sim
\big\vert{R_bT_2}\text{e}^{i\phi_2}\text{e}^{i\varphi_2}\left(\gamma_1T_aT_1\text{e}^{i\phi_1}\text{e}^{i\varphi_1}
+\gamma_2R_a\text{e}^{i\phi_3}\right)+\gamma_3T_b\big\vert^2
\nonumber\\
+\vert{R_bR_2}\vert^2\vert\gamma_1T_aT_1\text{e}^{i\phi_1}\text{e}^{i\varphi_1}+\gamma_2R_a\text{e}^{i\phi_3}\vert^2
+\gamma_1^2\vert{T_aR_bR_1}\vert^2\quad
\end{eqnarray}
The condition given by equation
\eqref{eq:condition_for_object2_invisible} for making object $2$
invisible becomes in this case
\begin{eqnarray}
\label{eq:suppressed_PDC_condition_GENERAL}
\gamma_1T_aT_1\text{e}^{i\phi_1}\text{e}^{i\varphi_1}+\gamma_2R_a\text{e}^{i\phi_3}=0
\end{eqnarray}
It is obvious now that the delays bring no new elements in our
condition: the phase shifts $\phi_1$ and $\phi_3$ can be grouped
together and eventually absorbed into $\varphi_1$. Moreover,
$\phi_2$ is useless in respect with any invisibility condition
regarding object $2$. If condition
\eqref{eq:suppressed_PDC_condition_GENERAL} is satisfied, the
single photo-count probability at $D_B$ is
\begin{eqnarray}
P_{B}\sim\gamma_1^2\vert{T_aR_bR_1}\vert^2+\gamma_3^2\vert{T_b}\vert^2
\end{eqnarray}
The coincidence counts at the detectors $D_B$ and $D_I$ can be
equally computed yielding $P_{B-I}\sim
\gamma_3^2\vert{T_b}\vert^2$.
It is not difficult to show that the condition for cancelling the
dependence of the coincidence counts at $D_A$ and $D_I$ on $T_2$
(i.e.
$\gamma_1R_aT_1\text{e}^{i\phi_1}\text{e}^{i\varphi_1}+\gamma_2T_a\text{e}^{i\phi_3}=0$)
cannot be simultaneously satisfied with equation
\eqref{eq:suppressed_PDC_condition_GENERAL}.

%

\end{document}